# Hi-M imaging of chromatin architecture in adult *Drosophila* brain cryosections

Christel Elkhoury Youhanna, Julie Garona, Marie Schaeffer, Christophe Houbron, Jean-Bernard Fiche, Olivier Messina, Marcelo Nollmann

*Centre de Biologie Structurale, Univ Montpellier, CNRS UMR5048, INSERM U1054, 29 rue de Navacelles, 34090 Montpellier, France*

Corresponding author: Marcelo Nollmann, marcelo.nollmann@cnrs.fr

**Abstract**

Hi-M combines fluorescence in situ hybridization (FISH), automated microfluidics, sequential imaging, and computational chromatin tracing to measure the three-dimensional organization of selected genomic regions in single cells. This chapter describes a Hi-M workflow adapted for cryosections of adult *Drosophila melanogaster* brains, enabling chromatin tracing while preserving tissue architecture and cell identity. The protocol covers Oligopaint library design and amplification, fixation, brain dissection, cryoprotection, cryosectioning, sequential RNA-FISH for cell-type identification, sequential DNA-FISH labeling, automated acquisition, and chromatin trace reconstruction. We also provide practical guidance for experimental design, sample preparation, tissue handling, image acquisition, and data analysis, highlighting critical steps that influence tissue integrity, hybridization efficiency, image registration, barcode detection, and chromatin trace reconstruction. The workflow is readily adaptable to different genomic loci and cell types, providing a robust approach for studying 3D genome organization in intact adult tissues.

# 1 Introduction

Chromatin folding contributes to transcriptional regulation by shaping the spatial organization of promoters, enhancers, and other regulatory elements [1,2]. Population-based chromosome conformation capture methods have revealed general principles of genome organization [3], yet they average chromatin contacts across many cells, obscuring cell-to-cell variability [4]. Imaging-based chromatin tracing methods complement chromosome conformation capture approaches by directly measuring three-dimensional (3D) chromatin organization in individual cells while preserving tissue architecture and cell identity [5–9].

Hi-M is a chromatin tracing method that combines RNA- and DNA-FISH with sequential imaging to visualize 3D chromatin organization and transcription in single cells while preserving tissue architecture [8,10] (**Figure 1a**). In a typical experiment, RNA-FISH is first performed using oligonucleotide libraries in which each target transcript is assigned a unique readout sequence. Fluorescent readout probes complementary to these sequences are then sequentially hybridized, imaged, and chemically removed over multiple imaging cycles, allowing the expression of many RNA species to be measured in the same sample (**Figure 1b, left**). Following RNA imaging, the RNA is degraded and a primary Oligopaint DNA-FISH library [11] targeting a defined genomic region is hybridized. The target region is subdivided into multiple genomic intervals (or barcodes), each assigned a unique readout sequence. Sequential hybridization of fluorescent readout probes is then repeated to determine the 3D positions of all barcodes and reconstruct chromatin organization with nanometer precision (**Figure 1b, right**). Fiducial markers are imaged throughout the experiment to align successive imaging cycles and register the RNA- and DNA-FISH datasets. Multiple regions of interest (ROI) are acquired for each experiment to increase throughput.

This chapter describes a Hi-M protocol for imaging chromatin organization in cryosections of adult *Drosophila* brains. Compared with previous Hi-M workflows developed for *Drosophila* embryos [12], the protocol introduces several key adaptations. Adult heads are fixed prior to brain dissection, brains are cryoprotected in sucrose and embedded in OCT, and thin cryosections are mounted on activated silane-coated coverslips. Cell types are identified by sequential RNA-FISH detection of marker transcripts, allowing chromatin traces to be assigned to defined cell types. The protocol is readily adaptable to different genomic regions and cell types.

The protocol is organized around eight sections: Design and amplification of RNA-/DNA-FISH libraries, fly brain fixation and cryosectioning, RNA-/DNA-FISH labeling, automated acquisition and analysis. The Notes section collects practical considerations and troubleshooting points that are often not visible in standard methods descriptions.

## 2 Materials

### 2.1 Reagents and Solutions

- 20X Saline-sodium citrate buffer (SSC) containing 3 M NaCl, 0.3 M sodium citrate (Thermo Fisher Scientific, cat. no. AM 9770)
- 16% Formaldehyde solution (w/v) (ThermoFisher Scientific, cat. no. 28906)
- 4′,6-Diamidine-2′-phenylindole dihydrochloride (DAPI; Roche, cat. no. 10236276001)
- Acetone (Merck, cat. no. 1000122500)
- Agarose Standard DNA Grade (Euromedex, cat. no. D5-E)
- AlexaFluor 488 readout probes used for fiducial (IDT DNA)
- Alexa-647 readout probes containing a cleavable disulfide bond. (IDT DNA)
- Ammonium Persulfate (APS; Fisher Scientific, cat. no. 17874)
- Ammonium acetate 5 M (Fisher Scientific, cat. no. 10534645)
- BSA (Roche, cat. no. 10711454001)
- Cetyl PEG/PPG-10/1 dimethicone. (ABIL EM-90, Evonik)
- D(+) Glucose anhydrous (Euromedex, cat. no. UG3050)
- DNA Clean & Concentrator kit - 100 µg capacity (Zymo, cat. no. D4029)
- DNA Clean & Concentrator kit - 25 µg capacity (Zymo, cat. no. D4033)
- Dextran sulfate (Sigma-Aldrich, cat. no. D8906)
- Diethyl ether (Sigma-Aldrich, cat. no. 296082)
- Dulbecco's phosphate-buffered saline (PBS) (Gibco, cat. no. 14190169)
- Ethyl acetate (Sigma-Aldrich, cat. no. 270989)
- GeneRuler 100 bp DNA Ladder (Fisher Scientific, cat. no. SM0243)
- Glycogen 5 mg/mL (Ambion, cat. no. AM9510)
- Glucose oxidase (Sigma-Aldrich, cat. no. G2133)
- HiScribe T7 High Yield RNA Synthesis Kit (New England Biolabs, cat. no. E2040S)
- KAPA Taq Kit with dNTPs (CliniSciences, cat. no. BK1003)
- Low Range ssRNA Ladder (New England Biolabs, cat. no. N0364S)
- Maxima H Minus Reverse transcriptase kit (Fisher Scientific, cat. no. 1324315)
- Methanol (Fisher Chemical, cat. no. A412-4)
- Mineral oil (500 mL; Sigma-Aldrich, cat. no. M5904)
- Oligo clean & concentrator kit (Zymo, cat. no. D4060)
- Oligo binding buffer (Zymo, cat. no. D4060-1-40)
- Poly-L-lysine solution (Sigma-Aldrich, cat. no. P8920)
- RNase A (Sigma-Aldrich, cat. no. R6513)
- RNA Loading Dye 2X (New England Biolabs, cat.no. B0363S)

- RNA Ribonuclease Inhibitors (Promega, cat. no. N2515)
- SYBR Gold Nucleic Acid Gel Stain (Fisher Scientific, cat. no. S11494)
- SYBR Safe nucleic acid gel stain (Invitrogen, cat. no. S33102)
- TAE Buffer 50X (Tris-acetate-EDTA) Thermo Scientific, cat. no. 10399519)
- TCEP (tris(2-carboxyethyl)phosphine) ) (UBP Bio, cat. no. P1021)
- TEMED (Thermo Scientific, cat. no. 17919)
- Tris Base, BM grade (Euromedex, cat. no. 200923-A)
- Triton X-100 (250 mL; Sigma-Aldrich, cat. no. T8787)
- Tween-20 (500 mL; Sigma-Aldrich, cat. no. P2287)
- dNTP Set 100 mM Solution (Fisher Scientific, cat. no. 10083252)
- deionized formamide (100 ml; Amresco, cat. no. 0606)
- Boric acid (Sigma, B0394-100G)
- Urea (Euromedex, cat. no. EU0014-B)
- TEMED (Biosolve BV, cat. no. 20102399)
- Ethanol (Sigma aldrich, cat. no. 8.18760.2500)
- 6X DNA loading Buffer
- Neg-50 Blue (Epredia, cat. no. 6502B)
- Marabu Fixogum Cement Rubber Tube Stencil

**2.2 Equipment**

- PCR Machine (Bio-Rad, T100 Thermal cycler)
- Positive displacement micropipette Gilson Microman M250 (Gilson, cat .no FD10005)
- 1.8 mL plastic cryotube (ThermoFisher Scientific, cat. no 377267)
- Magnetic stirring bar (BelArt, cat. no. 371191083)
- NanoDrop spectrophotometer (Thermo Scientific, model no. ND-1000UV/Vis)
- Vortex, standard mini vortex (VWR )
- Falcon 15-mL Conical Centrifuge Tubes (Fisher Scientific, cat. no. 14-959-53A)
- Falcon 50-mL Conical Centrifuge Tubes (Fisher Scientific, cat. no. 14-959-49A)
- Tabletop centrifuge (Eppendorf, cat. no. 5424)
- Dry Bath (Fisher scientific, FB15103)
- Syringe 30 mL (Terumo, cat. no. SS-30S)
- Syringe 20 mL (Terumo, cat. no. SS-20S2)
- Cellulose Nitrate Filter (Sartorius Stedim Biotech, cat. No. 11306--47------N )
- Shandon Base Mold (Thermo scientific, cat. No. 41741)
- 30 mm 0.22 µM Syringe Filter (Clear Line, cat. No. 146560)
- Water bath Grant Instruments JBN5 (Fisher Scientific, cat. no. 15177015)
- Thermomixer-AccuTherm Microtube Shaking Incubator (Labnet, cat. no. I-4001-HCS)
- Disposable scalpel (Swann-Morton, cat. no. 0516)
- Plastic petri dishes (Greiner Bio-One; 60 mm diameter petri dish, cat. no.628163)

- Microscope coverslips (Bioptechs Inc., cat. no. 40-1313-0319)
- Microfluidics FCS2 chamber, no heat and low dead volume (Bioptechs Inc., cat. no. 060319-2-03 )
- Adult Drosophila culture equipment and CO2 anesthesia setup (Flystuff, FLY1032, Scientific Laboratory Supplies).
- Dissection microscope (ZEISS Stemi 305 LAB Stereo Microscope – no. 435063-9020-100)
- Rotating wheel (GRANT Instruments, PTR-35)
- Mini orbital shaker (Fisher scientific, cat. no. 10309644 )
- Water bath (GRANT Instruments, JB Nova 5)
- Dumont #55 Forceps (Fine Science Tool, cat. no. #11295-51)
- Cryostat suitable for 10 micrometer brain sections (Thermo scientific, Cryostar NX50)
- Home-made widefield epifluorescence microscope with 405 nm, 488 nm, 641/642 nm excitation laser lines, and 785 nm autofocus lasers; 60X 1.2 NA water immersion objective; piezo objective stage; motorized XY stage; sCMOS camera; focus stabilization system. A full description of this microscope was provided earlier [12].

**2.3 Reagent setup**

- PCR oil phase: 95.95% mineral oil, 4% ABIL EM-90, and 0.05% Triton X-100 (v/v/v). If you have access to a positive displacement pipette, you can conveniently pipette 2 mL of ABIL EM90, 25 µL of Triton X-100, and 47.975 mL of mineral oil into a 50-mL Falcon tube. When adding the mineral oil, do so in two steps, vortexing in between. In case a positive displacement pipette is not available, you can accurately measure the volume by weight. To prepare 50 mL of the PCR oil phase, weigh 20.3 g of mineral oil (approximately 24 mL) directly into a 50 mL Falcon tube. Then, add 1.86 g of ABIL EM90 and 0.0265 g of Triton X-100, vortex the mixture thoroughly, and allow it to sit for 5 minutes. Next, add 20 g of mineral oil and homogenize the mixture by gently inverting the tube. Finally, create 20 mL aliquots of the PCR oil phase and store them at 4°C indefinitely.
- Water-saturated diethyl ether : 3 mL of diethyl ether, 3 mL of ddH$_2$O. Vortex for 30 sec. Allow the mixture to settle and use the organic upper phase. Prepare freshly.
- Water-saturated ethyl acetate : 2 mL of ethyl acetate, 2 mL of ddH$_2$O. Vortex for 30 sec. Allow the mixture to settle and use the organic upper phase. Prepare freshly.
- 4% (w/v) formaldehyde in PBS (8mL): 2 mL of 16% formaldehyde (w/v), 6 mL of PBS. The solution can be stored at -20°C for several months.
- TBE 10X solution (500 mL) : 60.55 g of Tris base, 30.9 g Boric acid and 3.7 g of EDTA. Adjust volume to 500 mL with ddH$_2$O. Store at RT indefinitely.
- Gel for denaturing urea polyacrylamide gel electrophoresis (PAGE). Mix 6 g urea, 1.25 mL TBE 10X and 3.5 mL of ddH$_2$O. Heat the solution at 60°C in a water bath until the urea is dissolved. Add 3.125 mL of acrylamide/bisacrylamide, 75 µL of APS 10% and 15 µL of TEMED. Cast the polyacrylamide gel in 0.75 mm thick spacers. Prepare freshly.
- 2X SSC (100 mL) : Dilute 10 mL of 20X SSC in 90 mL dH$_2$O. Filter the solution through a 0.22 µm membrane filter before use.
- 70 % Ethanol in 2X SSC (2 mL) : Add 600 µL of 2X SSC to 1.4 mL of EtOH 100%.

- 0.5% PBS Triton (PBST) : Add 250 µL of Triton X-100 to 49.75 mL of 1× PBS and mix thoroughly.
- 4% PFA in 1X PBS 0.5% Triton (2 mL) : Mix 500 µL of 16% PFA, 1500 µL 0.5% PBST. Prepare fresh.
- 30% Sucrose in 1X PBS (50 mL): Dissolve 15 g of sucrose in approximately 40 mL of 1× PBS, heat to 60°C until it dissolves and then adjust the final volume to 50 mL with 1× PBS. The solution can be stored at 4°C for several months.
- RNAse A solution: Dissolve 10 mg of RNAse A vial in 1 mL of $ddH_2O$. Make small aliquots and store at -20°C for up to a year.
- 50% (w/v) Dextran sulfate: Dissolve 25 g of dextran sulfate in 40 mL of $ddH_2O$. Heat to 37°C until it fully dissolves and then add $ddH_2O$ to a final volume of 50 mL. The solution can be stored at 4°C for several months.
- RNA-FISH Hybridization Buffer (HB): Mix 3 mL of formamide 100%, 1g of dextran sulfate, 1 mL 20X SSC, 500 µL tRNA stock (20 mg/mL), 100 µl of RVC stock (200 mM), 40ul RNAse free BSA (50 mg/mL), 5.3 mL of $ddH_2O$. Store at -20°C for up to several months.
- DNA-FISH Hybridization Solution (FHB): Mix 5 mL of formamide 100%, 2 mL of 50% (v/v) Dextran sulfate, 1 mL of 20X SSC, 500 µL of Salmon Sperm (10 mg/mL) and 1.5 mL of $ddH_2O$. Store at -20°C for up to several months.
- Sodium Citrate Buffer (10mM Sodium Citrate, 0.05% Tween 20, pH 6.0): Dissolve 2.94 g of tri-sodium citrate (dihydrate) in 1 L of $ddH_2O$. Adjust pH to 6.0 with 1N HCl, filter the solution through a 0.22 µm membrane filter and then add 0.5 ml of Tween 20 and mix well. Store this solution at room temperature for 3 months or at 4°C for longer storage.
- Wash Buffer 50% (WB50, 10 mL): Prepare by mixing 1 mL of 20× SSC and 5 mL of 100% formamide, then add 4 mL of $ddH_2O$ to obtain a final formamide concentration of 50% (v/v).
- Wash Buffer 40% (WB40, 100 mL): Prepare by mixing 10 mL of 20× SSC and 40 mL of 100% formamide, then adjust the final volume to 100 mL with $ddH_2O$ to obtain a final formamide concentration of 40% (v/v). Filter the solution through a 0.22 µm membrane filter before use.
- Wash Buffer 30% (WB30, 100 mL): Prepare by mixing 10 mL of 20× SSC and 30 mL of 100% formamide, then adjust the final volume to 100 mL with $ddH_2O$ to obtain a final formamide concentration of 30% (v/v). Filter the solution through a 0.22 µm membrane filter before use.
- Wash Buffer 20% (WB20, 10 mL): Prepare by mixing 1 mL of 20× SSC and 2 mL of 100% formamide, then add 7 mL of $ddH_2O$ to obtain a final formamide concentration of 20% (v/v).
- Wash Buffer 10% (WB10, 10 mL): Prepare by mixing 1 mL of 20× SSC and 1 mL of 100% formamide, then add 8 mL of $ddH_2O$ to obtain a final formamide concentration of 10% (v/v).
- 5X PBS buffer for Imaging buffer (100 mL): Dissolve 4.45 g of Na2HPO4.2H2O 250 mM, 612.4 mg of KH2PO4 45 mM , 4g of NaCl 685 mM, 100.6mg of KCl 13.5 mM in 90 mL of $ddH_2O$, then adjust the final volume to 10 mL with $ddH_2O$ to obtain a final volume of 100 mL. Adjust pH to 7.4. Filter solution through a 0.22 µm filter. The solution can be stored at -20°C for several months.
- 50% (w/v) Glucose (40 mL): Dissolve 20 g of glucose, 30 mL $ddH_2O$. Heat to 60°C until it dissolves and then add $ddH_2O$ to 40 mL and pass it through a 0.22 µm filter. The solution can be stored at RT for several months.
- 1 M NaCl solution (50 mL): Dissolve 2.92 g of NaCl, 30 mL of $ddH_2O$. Complete to 50 mL with $ddH_2O$ and pass it through a 0.22 µm filter. The solution can be stored at RT for several months.

- 1 M Tris-HCl pH=8 solution (50 mL): 6 g of Tris base, 30 mL of $ddH_2O$. Using a pH meter, slowly add HCl using a glass pasteur pipette to reach the desired pH. Complete to 50 mL with $ddH_2O$, and pass it through a 0.22 µm filter. The solution can be stored at RT for several months.
- 55 mM NaCl in 11 mM Tris-HCl pH=8 solution (50 mL): mix 2.75 mL of 1 M NaCl solution, 0.55 mL of 1M Tris-HCl pH=8 solution, and 46.7 mL of $ddH_2O$. Prepare freshly.
- Gloxy solution (1 mL): 50 mg Glucose oxidase, 100 µL catalase and 900 µL of 55 mM NaCl in 11 mM Tris-HCl pH=8. Make 60 µL aliquots and store at -20°C. Solution is stable for several months. Defrost on the day of the experiment and keep it on ice until use. If there is a precipitate, spin it down and employ the supernatant. Once defrozen, the aliquot should be used within one week.
- Imaging Buffer IB: Mix 1.1 mL of 50% (w/v) glucose, 9.9 mL of 1X PBS (diluted from 5X PBS) and 110 µL of Gloxy. Add Gloxy solution just before using the solution and mix. Once the tubing is introduced, add a layer of mineral oil to prevent contact with oxygen from the ambient.
- TCEP chemical bleaching solution (10 mL): Mix 0.5 mL of TCEP 1M, 9.5 mL of 2X SSC. Prepare the solution right before its use.

### 2.4 Software and Online Resources

- Oligopaint designs for DNA-FISH probes were designed using oligopaint-design [12], or ProbeDealer [13].
- RNA-FISH libraries were designed using OligoStan [14].
- Hi-M datasets were acquired using Qudi-HiM [15].
- Hi-M data were analyzed using the pyHiM analysis software package [16].
- DAPI images from DNA-FISH and RNA-FISH imaging experiments were registered using the *RNA_to_DNA_registration* module from pyHiM_tools. Instructions are provided in the *README_registration.md* file.
- Chromatin traces are assigned specific transcriptional states using the *trace_assign_mask.py* script of traceratops.
- RNA-FISH 3D image stacks are manually shifted using the *image_shift.py* script of pyHiM_tools. Instructions are provided in the documentation page.

## 3 Methods

### 3.1 Design of DNA/RNA-FISH libraries

DNA- and RNA-FISH libraries are designed using different computational pipelines (see below), but share a common primary probe architecture. In both cases, a sequence complementary to the target genomic region or transcript is flanked by upstream and downstream readout sequences that are unique to each DNA-FISH barcode or RNA target (**Figure 2a**). RNA libraries do not require PCR amplification, and therefore do not contain universal primers. Two strategies can be used to image readout sequences. In direct labeling, a fluorescently-labeled readout oligonucleotide hybridizes directly to the readout sequence. This approach is straightforward and can provide higher specificity, but requires one fluorescently-labeled oligonucleotide for each imaging cycle, increasing synthesis costs. In indirect labeling, a non-fluorescent ‘bridge’ oligonucleotide first hybridizes to the readout sequence and contains an additional binding site for a fluorescently-labeled ‘readout probe’. This strategy also requires one bridge oligonucleotide per imaging cycle, but the same fluorescent readout probe can be reused throughout the experiment, making it considerably more economical.

DNA-FISH library design:

1. Clone the oligopaint-design repository.
2. Choose the genomic region to be traced and define the desired number of barcodes, barcode size, and minimum number of primary probes per barcode. We recommend a minimum of 45 oligonucleotides per barcode [17].
3. In the *input_parameters.json* file, enter the various parameters required, such as the chromosome of interest, the location of the .bed files, the start-end coordinates of the target region, the desired number of loci, their genomic size, and the minimum number of primary probes per locus.
4. The *Library_Design.py* script will: i) calculate the genomic coordinates for each barcode (start, end), ii) select primary probe sequences for each locus, iii) concatenate primary sequences with readout sequences and with universal primers, iiii) check the homogeneity of the size of the differents barcodes.
5. Several output text files are created after running *Library_Design.py*. *Library_Summary.csv* contains a table summarizing the information for each barcode (number, start-end position, readout probe, primer forward, primer reverse, number of probes). *outputParameters.json* is a JSON dictionary file containing the parameters used to generate the library. *Full_sequence_only.txt* contains the raw primary probe sequences for the oligos used to order the microarray.
6. Order the microarray from an oligopaint synthesizer company (e.g. Genscript). If multiple libraries will be pooled, assign non-overlapping universal primer pairs so that each library can be selectively amplified.

RNA-FISH library design:

7. Clone the oligostan-python repository.
8. Create a fasta-formatted input file containing the sequence against which you want to design RNA probes (e.g. exon, 5’-UTR, intron).
9. Run oligostan-python using the graphical-user interface.
10. The list of output oligonucleotides will be generated in the Probes_your_sequence folder located where you had your fasta-formatted input file.
11. Manual or automated steps are then used to curate this list of oligonucleotides to prevent off-target binding.
12. Readout sequences are appended to the 3’ and 5’-ends of each selected oligonucleotide. Different readout sequences are added for different RNA species.
13. Order oligonucleotides using the oligopool service (Integrated DNA Technologies).

### 3.2 Amplification of Oligopaint Libraries

Emulsion PCR

Perform the following steps in a 4°C cold room. Transfer all reagents and equipment to the cold room well in advance to allow them to equilibrate to 4°C before use.

14. Perform emulsion PCR to amplify the starting Oligopaint libraries, which can be limiting, in a non-biased manner. For this, set up a PCR Master Mix for each library as indicated in Table 1 and keep it on ice until needed.
15. Pre-chill a 1.8 ml cryotube in the freezer with a stirring bar, place it on the center of a controlled stirring plate.
16. Transfer 600 µL of PCR oil phase to the tube with a positive displacement pipette. Stir at 1000 rpm for at least 1 minute.
17. While the stirring bar is still spinning, add 100 µL of the emulsion PCR master mix in steps of 20 µL increments using a P20 pipette (i.e. dispense 20 µL 5 times).
18. Stir at 1000 rpm for 10 minutes, the emulsion should appear milky white and foamy.
19. Transfer the emulsion to a PCR strip tube (~8 x 75 µL) with a positive displacement pipette.

Forward primer is the 5’ => 3’ whereas reverse primer is the reverse complement of the reverse universal priming sequence.

20. Perform the PCR using PCR program 1 in Table 2. PCR products can be stored at 4°C for a few days.

Emulsion PCR breaking

21. Pool the emulsion PCR reactions in a 1.5 mL eppendorf tube.
22. Add 1 µL of 6X DNA loading buffer to visualize the aqueous phase.
23. Add 200 µL of mineral oil and vortex for 30 sec.
24. Centrifuge for 10 min at 12000G and remove the upper organic phase.
25. Add 1 mL of water-saturated diethyl ether and vortex for 1 min. Centrifuge at 12000G for 1 min and remove the diethyl ether upper phase.
26. Add 1 mL of water-saturated ethyl acetate and vortex 1 for min. Centrifuge at 12000G for 1 min and remove the ethyl acetate upper phase.
27. Add 1 mL of water-saturated diethyl ether and vortex for 1 min. Centrifuge at 12000G for 1 min and remove the diethyl ether upper phase.
28. Evaporate the residual diethyl ether by incubating the tube at 37°C for 5 min with the cap open. The final volume should be around 80 µL. PCR products can be stored at 4°C for a few days.

DNA Purification

Purify the DNA by using Zymo Oligo Clean & Concentrator kit.

29. Mix 80 µL of DNA from the emulsion PCR breaking, 160 µL of oligo binding buffer and 320 µL of ethanol. Homogenize the solution by pipetting up and down 10 times.
30. Follow the manufacturer's instructions up to the DNA elution.
31. Repeat elution with an extra 15 µL of water.
32. Quantify DNA concentration using a NanoDrop by directly taking 2 µL of purified PCR product. Concentration should be between 20-40 ng/µL.
33. Run a gel electrophoresis to check for a single band amplification with 200 ng of PCR product in a 1.5% agarose TAE gel with 0.01% SYBR Safe at 100 V for 45 min. Purified products can be frozen at -20°C for several months.

Small-scale limited-cycle PCR

The limited number of cycles is performed to find the cycle number where the PCR is still at its exponential phase. Perform this step before proceeding to the large-scale PCR. The T7 promoter sequence (5'-TAATACGACTCACTATAGGGT-3') should be added to the reverse primer used for the emulsion PCR step to allow for the reverse transcription step.

34. Perform the small scale limited-cycle PCR by setting up the following reaction mix for 8 tubes as indicated in Table 3.
35. Run the PCR program 2 in Table 4. At the end of the extension step, remove one PCR tube after each cycle from cycles 8 to 15. To do so, briefly open the thermocycler immediately after the extension phase, remove the corresponding tube, place it on ice, close the lid, and promptly resume the PCR program. PCR products can be left overnight (ON) at 4°C or frozen for up to a month at -20°C.
36. Run 20 µL of the PCR product with 6X DNA loading dye in a 1.5% agarose gel with 0.01% SYBR Safe at 100 V for 45 min.
37. Find the cycle with a single band of the expected size and the maximum intensity (Fig. 2c).

Large-scale limited-cycle PCR

This step will generate a big quantity of Oligopaints. The T7 promoter sequence should be added to the reverse primer to allow for the reverse transcription step.

38. Prepare a reaction mix for 16 tubes as indicated in Table 3
39. Split the volume of the mix in Table 3 into 16 x 50 µL PCR tubes.
40. Run the PCR program 2 from Table 4 using the previously optimized number of cycles. Add a last extension cycle of 5 min at 72°C. PCR products can be safely stored for months at -20°C.
41. Run 20 µL of the PCR product in an agarose gel as in step 36 to check that the PCR was successful.

PCR product purification

For this step we use a Zymo DNA Clean & Concentrator kit with 25 µg capacity.

42. Collect the 50 µL-aliquots from the previous step in a 15 mL falcon tube.

43. Proceed to DNA column purification according to the manufacturer's instructions. Elute using 30 µL of DNAse- and RNAse-free water.
44. Quantify product concentration with a NanoDrop using double-stranded DNA parameters.
    This typically requires a 1/10 dilution of a 2 µL aliquot of the purified product. Concentration should be between 30-50 ng/µL.
45. Run the remainder of the 1/10 stock dilution in a 1.5% agarose gel as previously described in step 36. Check for a single band of the expected size (Figure 2d).

*In vitro* transcription

This step is a high-yield reaction that further amplifies the template molecules as well as converts them into RNA. It is necessary to keep RNAse-free conditions at all times.

46. Prepare the in vitro transcription reaction mix as indicated in Table 5.
47. Split the volume into 3x20 µL PCR tubes and incubate at 37°C for 12-16 h in a thermocycler. In vitro transcription products can be frozen for months at -80°C.
48. Take 5 µL and purify with a Zymo Oligo Clean & Concentrator kit,10 µg capacity, according to manufacturer's instructions, using 15 µL of DNAse and RNAse-free water to elute purified product. The purification is only performed with a small aliquot to control if the in vitro transcription was successful and to estimate the RNA concentration in the non-purified RNA solution.
49. Make a 1/10 dilution to perform a quantification of the purified RNA on NanoDrop using RNA parameters. Concentration should be between 0.5-2 µg/µL. The concentration obtained allows us to estimate concentration in non-purified RNA. For example, a 2 µg/µL concentration in the purified RNA can be translated to an estimated concentration of 6 µg/µL in the non-purified RNA considering a factor 3 dilution (from a 5 µL aliquot to a final volume of 15 µL). The total yield of the in vitro transcription step should be around 150-450 µg from a single transcription step (60 µL in total).

RNA quality check by urea PAGE

This step is realised to check the RNA quality by urea PAGE (**Figure 2e**).

50. Perform a pre-run for 30 min in 1X TBE at 190 V to eliminate the excess of persulfate.
51. When finished, wash the wells with the running buffer. Load 100 ng of purified RNA per lane. Heat the samples at 95°C for 5 min and put it immediately on ice for 2 min. Perform the PAGE for 1 h at 190 V. For gel staining, incubate protected from light for 20 min in 30 mL of 1X TBE and 3 µL of SyBR Gold.

Reverse Transcription

52. Perform the reverse transcription reaction according to Maxima H Reverse Transcriptase kit by setting up the reaction mix indicated in Table 6. In this step, the non-purified RNA from in vitro transcription is directly used. The primer sequence is the same as for forward primers used in emulsion PCR or limited-cycle PCR.

53. Split the volume obtained in the previous step into two 1.5 mL tubes and incubate for 3 h at 50°C in a water bath. Reverse transcription products can be frozen for months at -20°C.

Alkaline Lysis

54. Perform RNA degradation by adding into each tube 300 µL of 0.5 M EDTA and 300 µL of 1 M NaOH and incubating at 95°C for 15 min in a water bath. This step allows to selectively degrade the RNA while keeping single stranded DNA.
55. Take a 10 µL aliquot to control for DNA concentration and to perform a gel electrophoresis as in step 36.

Purification of single-stranded DNA Oligopaints

56. Mix the 2 aliquots in a sterile 50-mL Falcon tube. Add 4.8 mL of oligo binding buffer and 19.2 mL of ethanol 100%. Homogenize and evenly distribute across two columns. Follow the manufacturer's instructions from this point on. Use the Zymo DNA Clean & Concentrator kit with 100 µg capacity. Elute each column with 150 µL of elution buffer.
57. Take a 10 µL aliquot to measure DNA concentration and to perform a gel electrophoresis as in step 36.
58. Pool the eluates to obtain a final volume of 300 µL and perform ethanol precipitation by directly adding to DNA elution, 24 µL of 5 M ammonium acetate, 6 µL of glycogen and 750 µL of 100% (v/v) ethanol that is kept at -20°C. Vortex and incubate 1 h at -80°C or overnight at -20°C.
59. Centrifuge at 13,000G for 1 h at 4°C. Discard the supernatant.
60. Wash the pellet with 1 mL of ice-cold 70% ethanol (v/v).
61. Centrifuge at 13,000G for 15 min at 4°C. Discard the supernatant. Allow the residual ethanol to evaporate by leaving the tube open at 42°C for 5 min.
62. Add 20 µL of DNase- and RNase-free water.
63. Resuspend the single stranded DNA (ssDNA) for 10 min at 37°C with agitation. After resuspension, keep on ice.
64. Quantify oligo concentration with a NanoDrop using ssDNA parameters. The total quantity of ssDNA should be in the order of 80-120 µg.
65. Control the quality of ssDNA by Urea PAGE as in step 50 and 51 (Figure 2e). This step allows to verify RNA degradation and the efficacy of reverse transcription step. Probes can be stored at -20°C for months.

## 3.3 Adult fly brain fixation, dissection, and cryosectioning

66. Collect adult flies 2-5 days after eclosion from the desired crosses.
67. Anesthetize the flies with $CO_2$ and place them on a $CO_2$ pad.
68. Dewax the flies by immersing them sequentially for 2-5 s in each of the following solutions using forceps and a separate tube for each wash: 70% ethanol/2× SSC, followed by two washes in 2× SSC. Immediately transfer flies into a 2 mL tube containing cold 4% PFA/0.5% PBST for fixation. Place the tube on a rotating wheel at 18 rpm and fix the flies for 3 h at room temperature. Up to 15 flies can be processed per tube.

69. Wash the fixed flies twice in cold PBS.
70. Dissect adult brains on a Silgar polymer plate in ice-cold PBS. Keep dissected brains on the plate for no more than 20 min.
71. Post-fix dissected brains in cold 4% PFA/0.5% PBST in 2ml tubes for 20 min at room temperature on a rotating wheel at 18 rpm.
72. Wash brains three times for 20 min each in cold 0.5% PBST on rotating wheel. Remove solutions slowly to avoid aspirating brains.
73. Leave brains in approximately 200 µL of 0.5% PBST after the final wash.
74. Cryoprotect brains by adding 1.8 mL of filtered 30% sucrose/1X PBS solution overnight at 4°C.
75. Transfer brains into a square cryomold, align them, let them dry a bit before adding the OCT (Neg50 blue) embedding matrix, and freeze at -80°C.

Coverslip Coating

76. Clean coverslips with 70% ethanol.
77. Treat the coverslips with air plasma for 30 sec.
78. Place coverslips into glass petri dishes and coat them with 100ul of 3-aminopropyltrimethoxysilane and incubate them for 5 min at room temperature.
79. Add nuclease-free water to completely cover the silanized coverslips and incubate for 10 min.
80. Wash twice with nuclease-free water for 10 min under agitation.
81. Add with 0.5% glutaraldehyde/1X PBS and incubate at room temperature for 30 min.
82. Rinse well with nuclease-free water for 10 min then remove water and let it air-dry.
83. Coat the silanized coverslips with 0.1 mg/mL poly-L-lysine diluted in nuclease free 1X PBS for 1 h at room temperature.
84. Wash with nuclease-free water and then incubate them in fresh nuclease-free water for 1 h to overnight at room temperature.
85. Remove the water and allow the coverslips to air-dry.
86. Cut 10 µm cryosections and carefully transfer the sections onto the coated coverslips.
87. Place the coverslips in Petri dishes and allow the tissue sections to air-dry for 1-2 h at room temperature.
88. For storage, close the Petri dishes, seal them with Parafilm, and store them at −20°C.

### 3.4 RNA-FISH labeling of fly brain cryosections

DAY1

89. Remove freshly cut cryosections from -20°C and allow them to equilibrate at room temperature for 5 min before removing the parafilm seal to prevent condensation from forming.

90. Open the Petri dish and allow the sections to air-dry for 30 min to 1 h at room temperature.
91. Fix sample by adding 1 mL of 4% PFA /1X PBS and incubate for 10 min at room temperature.
92. Wash sample with 1X PBS for 5 min.
93. Slowly add cold 70% ethanol/RNase free water on the coverslip.
    Keep the 70% ethanol/RNase-free water solution at -20°C and transfer it to 4°C approximately 1 h before use.
94. Seal the petri dish with Parafilm and incubate at 4°C overnight (up to 72 h).

DAY2

95. Re-hydrate the sample in 2X SSC for 5 min at room temperature.
96. Incubate the coverslip in WB30 for at least 3 h in the oven at 37°C.
97. Prepare a total of 20 µL of probe hybridization mix by combining Hybridization Buffer (HB) with 1.6 µL of each Oligopaint pool (250 µM). Adjust the volume of HB to reach a final volume of 20 µL.

NOTE. The concentration of the library may vary depending on the targeted transcript.

98. Pipette 20 µL of the hybridization mix in the middle of a clean glass Petri dish. Invert the coverslip (tissue side down) and carefully place it onto the droplet, ensuring that no air bubbles are trapped and seal the sides of the coverslip with glue.
99. Incubate for 36 h at 37°C in a humidified incubator.

DAY4

100. Carefully remove glue from around the coverslip.
101. Add 2X SSC to completely cover the coverslip, and with a razor blade, gently lift one edge of the coverslip until it detaches from the glass surface.
102. Transfer the coverslip to a small Petri dish with the tissue facing upward.
103. Wash twice in WB30 for 30 min each in the oven at 45°C.
104. Wash in 2X SSC, then store at 4°C in 2X SSC until imaging.

### 3.5 Acquisition of sequential RNA-FISH images

105. Rinse microfluidics tubing first from the 50% ethanol with filtered water and then with filtered 2X SSC before mounting the sample.
106. Mount the coverslip carrying brain cryosections in the FCS2 chamber. Fill the chamber slowly with 2X SSC and confirm that there are no leaks or air bubbles.
107. Place the chamber on the microscope and verify sample stability by flowing buffer through the chamber before beginning the experiment.
108. Stain nuclei with 0.5-1 ug/mL DAPI in PBS for 5-10 min.

109. Using DAPI and tissue morphology, select 20-30 regions of interest (ROI). Each ROI typically measures 320 x 320 micrometers and contained one Drosophila brain section.

110. For each RNA-FISH imaging cycle, inject a 40 nM readout oligonucleotide probe prepared in 2 mL of WB30. Use the labeling strategy described in Table 7.

111. Calibrate autofocus using the 785 nm reflection system before starting sequential imaging. Define the reference plane so that the full tissue thickness is captured across ROIs.

112. Acquire 3D images using the 405 nm and the 647 nm excitation lasers at 250 nm steps across a total range of approximately 10 µm. Repeat this process for all ROIs selected.

113. Chemically bleach the Alexa Fluor 647 readout probe by injecting TCEP chemical bleaching solution then wash with 2X SSC for 10 min before the next hybridization cycle.

114. Repeat the hybridization, wash, imaging, bleaching, and wash sequence for each RNA-FISH library.

Unmounting coverslips after RNA-FISH imaging

After RNA-FISH imaging, do not remove the chamber from the support immediately.

115. Using a permanent marker, draw a reference line across both the chamber and the support. This line will allow the chamber to be repositioned in the same orientation during reassembly (Figure 3a).

116. Before separating the chamber from the support, identify: the first ROI and mark it with a square and the last ROI and mark it with a circle (**Figure 3a**).

117. Add a third reference mark ('X') at another location to create a unique triangular arrangement with the square and circle. This makes the orientation unambiguous during realignment.

118. Remove the chamber from the support and clean the immersion oil from the underside of the chamber with ethanol.

119. Transfer the orientation marks ('square', 'circle', and 'X') onto the experimental coverslip by tracing them from the fluidics coverslip so that both coverslips carry identical reference marks (**Figure 3b**).

120. Carefully separate the experimental coverslip, gasket, and fluidics coverslip. Set the marked fluidics coverslip aside for later use during the DNA-FISH experiment, and proceed with the DNA-FISH protocol using the experimental coverslip (**Figure 3c**).

## 3.6 DNA-FISH labeling of fly brain cryosections

DAY 1

121. Turn ON the water bath at 80°C.

122. Treat the sample with RNAse A at 200 µg/ml in PBS for 1 h at RT under agitation.

123. Wash with 1X PBS.

124. Wash with a 10 mM sodium citrate permeabilisation buffer for 5 min at room temperature.

125. Permeabilize the sample in a sodium citrate buffer for 25 min at 80°C in a water bath. Change the buffer kept at 80°C every 5 min (at least 5 times).

126. Let it cool for 1 h at room temperature.

127. Wash in 2X SSC for 5 min.

128. Incubate in WB50 for 2 h at room temperature.

129. Prepare a total of 22 µL of probe hybridization mix by combining 19 µL of FHB, 1 µL of 100 µM fiducial oligonucleotides, and 2 µL of the Oligopaint library (5-10 µg/µL).

130. Pipette 22 µL of the hybridization mix in the middle of a clean glass Petri dish. Invert the coverslip (tissue side down) and carefully place it onto the droplet, ensuring that no air bubbles are trapped and seal the sides of the coverslip with glue.

131. Incubate the Petrie dish in a water bath for 3 h at 45°C. Adjust the water level so that it contacts only the bottom of the Petri dish, avoiding immersion of the sides or lid.

132. Heat shock the sealed sample by placing the Petrie dish on the heating block of the dry bath at 85°C for 5 min.

133. Transfer the glass Petrie dish to a humidified incubator and incubate overnight at 37°C.

DAY 2

134. Remove the glue carefully with tweezers.

135. Add 2X SSC to completely cover the coverslip, and with a razor blade, gently lift one edge of the coverslip until it detaches from the glass surface. Transfer the coverslip to a small Petri dish with the tissue facing upward.

136. Wash twice for 1 h each with a WB50 preheated at 37°C, on an orbital shaker at 65 rpm at room temperature.

137. Wash sequentially in pre-heated 37°C solutions as follows: WB40 for 20min, WB30 for 20 min, WB20 for 20 min, and WB10 for 20 min on an orbital shaker at 65 rpm at room temperature.

138. Incubate for 20 min in 2X SSC.

139. Post-fix samples for 10 min in 4% PFA/1X PBS.

140. Wash three times in 2X SSC.

141. Store labeled samples at 4°C in 2X SSC until imaging.

Realignment of coverslips

For realignment, reverse the procedure described above as follows:

142. Reassemble the experimental coverslip, gasket, and fluidics coverslip, ensuring that all orientation marks are properly aligned (**Figure 3c**).

143. Close the chamber and remove the orientation marks from the experimental coverslip with ethanol (**Figure 3b**).

144. Reposition the chamber on the microscope support by aligning the reference lines drawn on the chamber and support. Navigate to the square marker, corresponding to the first ROI (**Figure 3a**).

145. Using the DAPI channel, scan the coverslip to relocate the previously imaged ROIs by comparing the live DAPI image with the reference DAPI images acquired during the RNA-FISH experiment. Once all ROIs have been identified, proceed with DNA-FISH imaging.

### 3.7 Acquisition of sequential DNA-FISH images

146. Once the ROIs have been relocated, proceed with fiducial, mask, and DAPI labeling for DNA-FISH imaging according to the labeling procedure described in Table 8.
147. For indirect mask labeling, first inject the 40 nM bridge oligonucleotide prepared in 2 mL of WB40 to label the entire Oligopaint library. Then inject the corresponding 40 nM fluorescent readout probe prepared in 2 mL of WB40 according to the labeling scheme described in Table 8.
148. Check the autofocus and select the starting point for the Z stack. Select the total number of planes using the same settings as before. Set the same offset and psd settings in Qudi-HiM as those used for RNA-FISH imaging.
149. Acquire the DAPI, fiducial, and mask channels using the 405 nm, 488 nm and 639 nm excitation lasers, respectively.
150. Chemically bleach the mask labeled with Alexa Fluor 647 readout probe by injecting TCEP chemical bleaching solution, then wash with 2X SSC for 10 min before the next hybridization cycle.
151. Set the injection procedure for the sequential imaging according to the labeling scheme described in Table 9
152. Acquire the images of readout probe and fiducial channels respectively in 488nm and 645nm Repeat the same steps for each barcode in the oligopaint library.
153. After the experiment, rinse the fluid lines with water and then with 50% ethanol for storage.

### 3.8 Image processing and chromatin trace analysis

Image analysis consists of several main steps, including RNA/DNA realignment, assembly of inputs, pre-processing, detection of RNA/DNA-FISH, chromatin tracing and trace classification (**Figure 4**). These steps will be described in detail in the following sections.

#### Registration of RNA-FISH and DNA-FISH images

For each RNA-FISH imaging cycle, a DAPI image was acquired and aligned to the DAPI reference image from the DNA-FISH experiment using a global transformation comprising rotation, translation, and, when required, independent scaling along the two image axes (**Figure 4a**). The resulting transformation was then applied to the corresponding RNA-FISH image, placing RNA and DNA signals in a common coordinate system and enabling DNA-FISH traces to be associated with RNA-expression levels and cell-type identity. This process is implemented in the *RNA_to_DNA_registration* module from pyHiM_tools. Full execution instructions are provided in the *README_registration.md* file.

#### Reorganize data

All image stacks and experimental metadata are organized into a single directory for analysis with pyHiM, an open-source software package developed for the analysis of multiplexed DNA-FISH imaging datasets [16] (**Figure 4b**). For a complete description of the software architecture and analysis workflow, installation instructions, tutorials, and descriptions of the available modules and parameters, readers are referred to Devos *et al.* [16].

Briefly, two types of input files are required to run a pyHiM analysis. The first is a configuration file (*parameters.json*), which contains all the information required for the execution of the different pyHiM modules. The second consists of 3D image stacks. Three image types are required: (i) RNA-/DNA-FISH images acquired at each hybridization cycle; (ii) fiducial marker images that are acquired simultaneously with each RNA-/DNA-FISH cycle. These fiducial images correspond to either DAPI (for RNA-FISH) or to satellite DNA regions (for DNA-FISH); (iii) for DNA-FISH, images of the regions labeled by the entire oligopaints library (called oligopaints mask) are acquired to perform chromatin tracing. These images are assembled into folders, with each folder containing all the images for one ROI.

**Pre-processing**

A number of pre-processing steps including deconvolution, image projection and registration are applied (**Figure 4c**). These steps require the installation of pyHiM, deconwolf, and segmentation tools (either fiji, cellpose or ilastik).

154. Install deconwolf [18] and deconvolve all images. Although pyHiM can process raw images, the use of deconvolved images is recommended to improve analysis.
155. Install pyHiM following the instructions at the official documentation page (https://pyhim.readthedocs.io/).
156. Install fiji, cellpose or ilastik following the official installation instructions.
157. Generate 2D projections of all acquired 3D image stacks, including DNA-FISH, RNA-FISH, DAPI, fiducial, and oligopaints mask images (pyHiM module: *project*). This is useful for rapid visual inspection of the quality of the acquired datasets.
158. Perform global image registration to correct the overall 3D rigid displacement between imaging cycles (pyHiM module: *register_global*). This step finds the x, y and z shifts that maximize the overlap between reference and target images.
159. For DNA-FISH exclusively, perform local registration to correct residual misalignments after global registration (pyHiM module: *register_local*). These misalignments can occur when the sample undergoes local deformations during the acquisition process.

**Detection**

160. Segment nuclei and oligopaint masks using pyHiM (module: *mask_3d*) (**Figure 4d**). This is performed internally using deep learning networks trained to detect nuclei or point-spread-functions.
161. Segment registered RNA-FISH images using cellpose [19], ilastik [20], or fiji [21]. These segmentations can be performed manually (e.g. using fiji), or using trained networks with cellpose or ilastik. The aim is to segment the regions containing the cells engaged in active transcription.
162. Localize DNA-FISH spots using pyHiM (module: localize_3d) (**Figure 4d**). This module performs automated segmentation using a deep learning model trained to detect point-spread-functions, measures spot properties for each region, and fits the intensity distribution using a 3D Gaussian function to accurately detect the center of each DNA-FISH spot. For each spot, pyHiM reports its 3D coordinates together with quantitative properties such as intensity, signal to noise ratio, and other quality control descriptors in a unified localizations table.

**Trace and classify**

163. Perform tracing by combining DNA-FISH spot localizations and oligopaint mask information (pyHiM module: *build_traces*) (**Figure 4e**). Two tracing strategies are available. The first relies on spatial clustering using a KDtree nearest-neighbor algorithm and is suitable for sparsely distributed traces. The second assigns DNA localizations directly to segmented oligopaints masks, allowing reconstruction of chromatin traces on an allele-by-allele basis.

164. RNA-FISH segmentation images are used to assign chromatin traces to individual cells and link them to their corresponding RNA expression profiles (**Figure 4e**). For this, use the *trace_assign_mask.py* script of traceratops.

## 4 Notes

165. Brain dissection is a major source of variability. Keep brains cold, minimize the time on the dissection plate, and avoid mechanical deformation during transfer.

166. Slow aspiration is essential during brain washes. Fixed brains are small and can be lost easily if solutions are removed quickly.

167. Make sure to use freshly dissected and cryosectioned samples for RNA-FISH experiments.

168. Activated and coated coverslips are critical for tissue retention during long formamide washes and multi-day microfluidic acquisition.

169. When performing DNA-FISH after RNA-FISH, ensure that the orientation marks used for realignment remain visible throughout the procedure. If any marks fade, redraw them before proceeding.

170. All solutions injected through the fluidics system (WB, 2× SSC, $H_2O$, and ethanol) should be pre-filtered before use.

171. Fiducial labeling must be strong and stable because cycle registration depends on it. Weak fiducial signals will compromise downstream image analyses steps.

172. Inspection of raw images during the first cycles is recommended. Diffraction-limited barcode spots and recognizable brain morphology are quick indicators of experiment quality.

173. For rare cell populations, acquire enough ROIs and biological replicates to compensate for lower trace counts.

174. First-neighbor distance distributions provide a useful sanity check for chromatin tracing at kilobase resolution; adjacent barcodes should generally be close in 3D space.

## 5 Competing Interests

The authors have no conflicts of interest to declare that are relevant to the content of this chapter.

## 6 Acknowledgements

This project was funded by the European Union's Horizon 2020 Research and Innovation Program (Grant ID 724429) (M.N.), by French National Research Agency grants ANR-24-CE12-2240-02, and ANR-23-CE12-0023-01. We acknowledge funding from grants from French National Research Agency under the France 2030 program, with the reference numbers ANR-24-EXCI-0001, ANR-24-EXCI-0002, ANR-24-EXCI-0003, ANR-24-EXCI-0004, ANR-24-EXCI-0005. This project was provided with computing (HPC) and storage resources by GENCI at CINES thanks to the grant allocation A0190316693 on the supercomputer Adastra GENOA/ MI250x partitions.

## 9 Tables

**Table 1**. Emulsion PCR master mix.

| Reagent | Quantity |
|---|---|
| ddH2O | 79 µL |
| 10X Taq Buffer A | 10 µL |
| BSA (10 µg/µL) | 5 µL |
| dNTPs (10 mM) | 2 µL |
| Forward primer (100 µM) | 1 µL |
| Reverse primer (100 µM) | 1 µL |
| Kapa Taq DNA polymerase (5 U/µL) | 1 µL |
| ssDNA library (10-30 ng/µL) | 1 µL |

**Table 2.** PCR program 1

| Cycle 1 | 95°C 2 min |
|---|---|
| Cycle 2-31 | 95°C 15 sec<br>60°C 15 sec<br>72°C 20 sec |
| Cycle 32 | 72°C 5 min |

**Table 3.** Small-scale limited-cycle PCR setup.

| Reagent | Quantity per tube |
|---|---|
| Kapa buffer A | 5 µL |
| dNTPs (10 mM) | 1 µL |
| Forward primer (100 µM) | 0.5 µL |
| Reverse primer with T7 promoter (100 µM) | 0.5 µL |
| Template emulsion PCR (1 ng/µL) | 2.5 µL |
| Kapa polymerase (5 U/µL) | 0.5 µL |
| ddH2O | 40 µL |

**Table 4.** PCR program 2

| Cycle 1 | 95°C 5 min |
|---|---|
| Cycle 2-15 | 95°C 30 sec<br>60°C 45 sec<br>72°C 30 sec |

**Table 5.** In vitro transcription solution mix.

| Reagent | Quantity |
|---|---|
| Purified PCR product | 6 µg template DNA |
| ATP (100 mM) | 6 µL |
| UTP (100 mM) | 6 µL |
| CTP (100 mM) | 6 µL |
| GTP (100 mM) | 6 µL |
| 10X T7 buffer | 6 µL |
| Rnase inhibitor (40 U/µL) | 2.25 µL |
| HiScribe T7 polymerase | 6 µL |
| ddH2O | Make up to a final volume of 60 µL |

**Table 6.** Reverse transcription mix.

| Reagent | Quantity |
|---|---|
| Non purified transcription product | 150 µg |
| dNTP mix (100 mM) | 12 µL |
| Forward Primer (100 µM) | 50 µL |
| 5X Maxima buffer | 240 µL |
| RNAsin Plus (40 U/µL) | 30 µL |
| Maxima H reverse transcriptase (200 U/µL) | 30 µL |
| ddH2O | Make up to a final volume of 1200 µL |

**Table 7.** Readout probe labeling and acquisition for RNA-FISH.

| Step | Solution | Volume (µL) | Flow Rate (µL/min) |
|---|---|---|---|
| 1 | Readout probe (645 nm) in WB30 | 1500 | 200 |
| 2 | Incubate for 10 min | | |
| 3 | WB30 | 2000 | 200 |
| 4 | 2X SSC | 1500 | 200 |
| 5 | DAPI | 1500 | 200 |
| 5 | Imaging buffer | 1000 | 200 |
| 6 | Image Acquisition in 405nm and 645nm | | |
| 7 | TCEP | 1000 | 200 |
| 8 | 2X SSC | 1000 | 200 |

**Table 8**. Fiducial and DAPI labelling for DNA-FISH.

| Step | Solution | Volume (µL) | Flow Rate (µL/min) |
|---|---|---|---|
| 1 | Bridge | 1500 | 200 |
| 2 | Incubate for 10 min | | |
| 3 | WB40 | 2000 | 200 |
| 4 | Readout probe for mask (S-S 645 nm) and Fiducial in (488 nm) | 1500 | 200 |
| 5 | Incubate for 10 min | | |
| 6 | WB40 | 2000 | 200 |
| 7 | 2X SSC | 1500 | 200 |
| 8 | DAPI | 1500 | 200 |
| 9 | Incubate for 5 min | | |
| 10 | 2X SSC | 1500 | 200 |
| 11 | Imaging buffer | 1000 | 200 |
| 12 | Image Acquisition in 405nm and 488nm | | |

**Table 9.** Readout probe labeling and acquisition for DNA-FISH.

| Step | Solution | Volume (µL) | Flow Rate (µL/min) |
|---|---|---|---|
| 1 | Readout probe (645 nm) | 1500 | 200 |
| 2 | Incubate for 10 min | | |
| 3 | WB40 | 2000 | 200 |
| 4 | 2X SSC | 1500 | 200 |
| 5 | Imaging buffer | 1000 | 200 |
| 6 | Image Acquisition in 488nm and 645nm | | |
| 7 | TCEP | 1000 | 200 |
| 8 | 2X SSC | 1000 | 200 |

## 8 Figures

**Figure 1. Overview of the Hi-M workflow.**

**a.** Schematic of the Hi-M workflow. Adult brain cryosections mounted in a microfluidic chamber are imaged by automated sequential FISH hybridization. Regions of interest (ROIs) are selected for imaging, allowing simultaneous analysis of RNA expression and 3D chromatin organization at single-cell resolution.

**b.** Sequential imaging strategy. RNA-FISH is first performed through multiple cycles of readout-probe hybridization, imaging, and chemical probe removal to identify cell types and measure transcript abundance. Nuclei are re-imaged at each cycle using DAPI-staining to register images. Following RNA imaging, DNA-FISH labeling is performed, the coverslip is realigned, and images of nuclei are used to align RNA and DNA channels. Finally, sequential DNA-FISH imaging is performed to reconstruct chromatin traces from individual genomic barcodes using fiducial-based image registration.

**Figure 2. RNA/DNA labelling strategies, Oligopaint amplification and sample preparation workflow.**

**a.** Schematic representation of primary Oligopaint and Oligopool probes showing the target-specific homology region, the readout binding region used for sequential imaging, and the universal priming regions in blue is present only in Oligopaint probes. Both probe libraries can be detected using either indirect labelling with a bridge oligonucleotide followed by a fluorescent readout probe, or direct labelling in which the fluorescent readout probe hybridizes directly to the probe readout sequence.

**b.** Amplification workflow for Oligopaint libraries, including PCR amplification, *in vitro* transcription, reverse transcription, alkaline RNA hydrolysis, and recovery of single-stranded DNA probes.

**c.** Example of an agarose gel electrophoresis of the small-scale limited-cycle PCR. L, DNA ladder. The expected 166-nt size is detected between the 100- and 200-nt ladder bands. A non-specific product (~300 nt) appears from cycles 11-14. Therefore, the amplification was stopped after 10 cycles.

**d.** Example of an agarose gel electrophoresis of the large-scale limited-cycle PCR. L, DNA ladder. Lane 1 corresponds to a PCR performed without a template and lane 2 with a template. dNTPs are observed at the bottom. Lane 3 corresponds to column-purified PCR products.

**e.** Example of an urea PAGE result. L, low-range ssRNA ladder. Lanes 1-4 show bands migrating close to the 150-nt ladder marker. Lane 1: 200 ng emulsion PCR break; lane 2: 200 ng RNA generated by in vitro transcription, exhibiting a higher apparent molecular weight due to the T7 promoter sequence; lanes 3 and 4: 200 ng ssDNA before and after precipitation, respectively.

**f.** Preparation of adult *Drosophila* brain cryosections for Hi-M imaging. Brains are dissected, cryoprotected, embedded in OCT, cryosectioned, and mounted onto activated silane-coated coverslips prior to RNA- and DNA-FISH labeling.

**Figure 3. Realignment of brain cryosections following sequential RNA- and DNA-FISH imaging.**

**a.** Following RNA FISH imaging, reference marks are added to the chamber and microscope support, and the marks were identified to preserve sample orientation.

**b.** After chamber disassembly, the orientation marks are transferred from the fluidics coverslip to the experimental coverslip.

**c.** The fluidics coverslip is preserved while the experimental coverslip undergoes DNA FISH labeling.

For DNA FISH imaging, the chamber is reassembled using the reference marks to restore the original orientation and relocate the previously imaged ROIs.

**Figure 4. Registration and analysis workflow for combined RNA- and DNA-FISH datasets.**

**a.** A rigid 3D transform is found and applied to realign RNA- and DNA-FISH image stacks based on nuclei (DAPI) images.

**b.** The inputs for the registration of RNA- and DNA-FISH datasets are assembled into a common folder where pyHiM is run.

**c.** Images are z-projected for visual inspection, and registered to remove drift between acquisition cycles.

**d.** Nuclei, Oligopaints masks and RNA-FISH images are segmented. DNA-FISH spots are localized.

**e.** Chromatin are traced and trace tables are built for each ROI. RNA-FISH segmentations are used to link chromatin traces to RNA expression patterns.

**a**

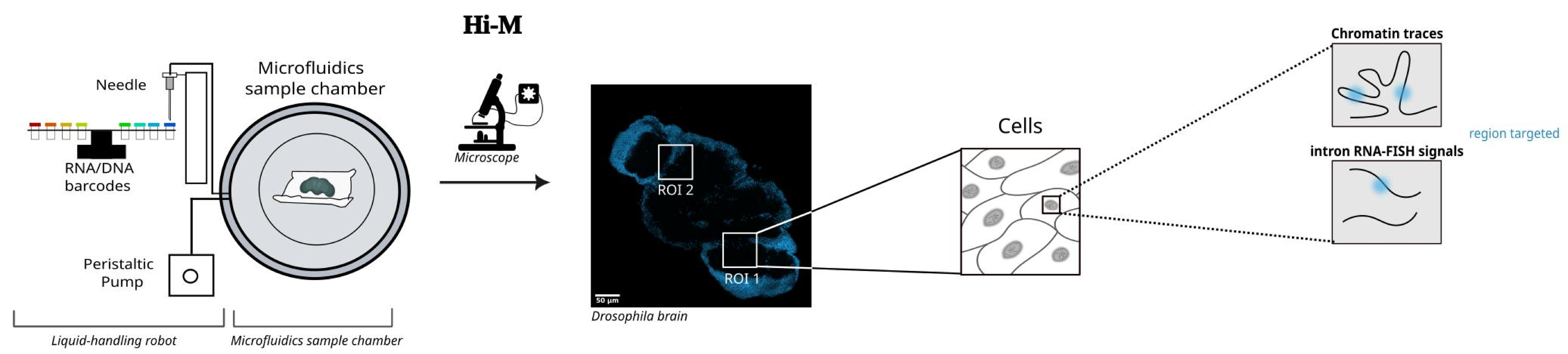

Hi-M
Needle
Microfluidics sample chamber
RNA/DNA barcodes
Peristaltic Pump
Liquid-handling robot
Microfluidics sample chamber
Microscope
ROI 2
ROI 1
Drosophila brain
Cells
Chromatin traces
region targeted
intron RNA-FISH signals

**b**

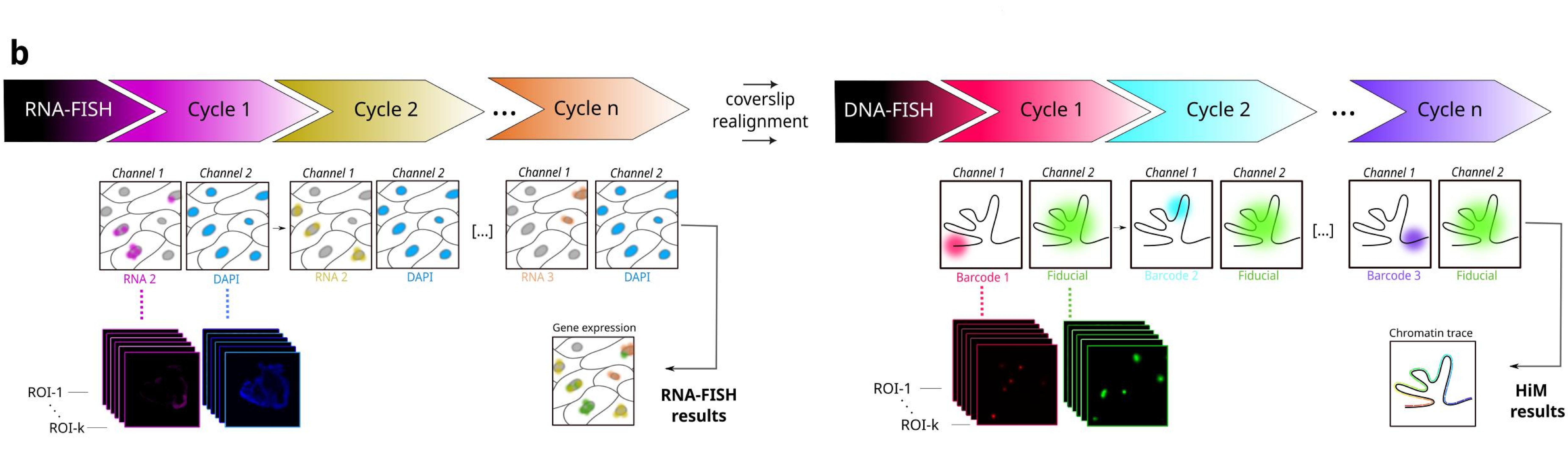

RNA-FISH
Cycle 1
Cycle 2
Cycle n
coverslip realignment
DNA-FISH
Cycle 1
Cycle 2
Cycle n
Channel 1
Channel 2
RNA 2
DAPI
RNA 3
[...]
ROI-1
ROI-k
Gene expression
RNA-FISH results
Barcode 1
Fiducial
Barcode 2
Barcode 3
Chromatin trace
HiM results

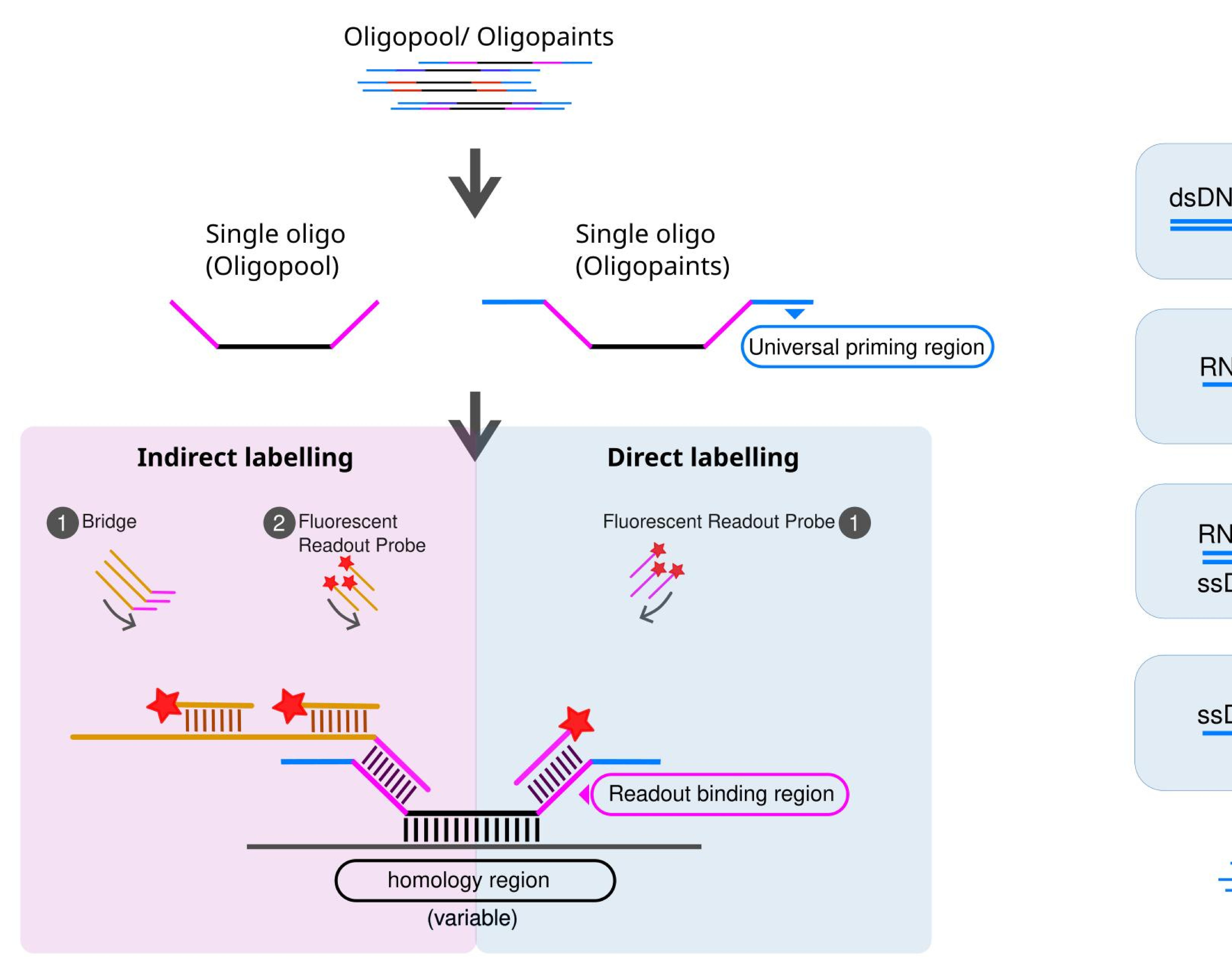

a. RNA/DNA labeling strategies
Oligopool/ Oligopaints
Single oligo
(Oligopool)
Single oligo
(Oligopaints)
Universal priming region
Indirect labelling
Direct labelling
1 Bridge
2 Fluorescent
Readout Probe
Fluorescent Readout Probe 1
Readout binding region
homology region
(variable)

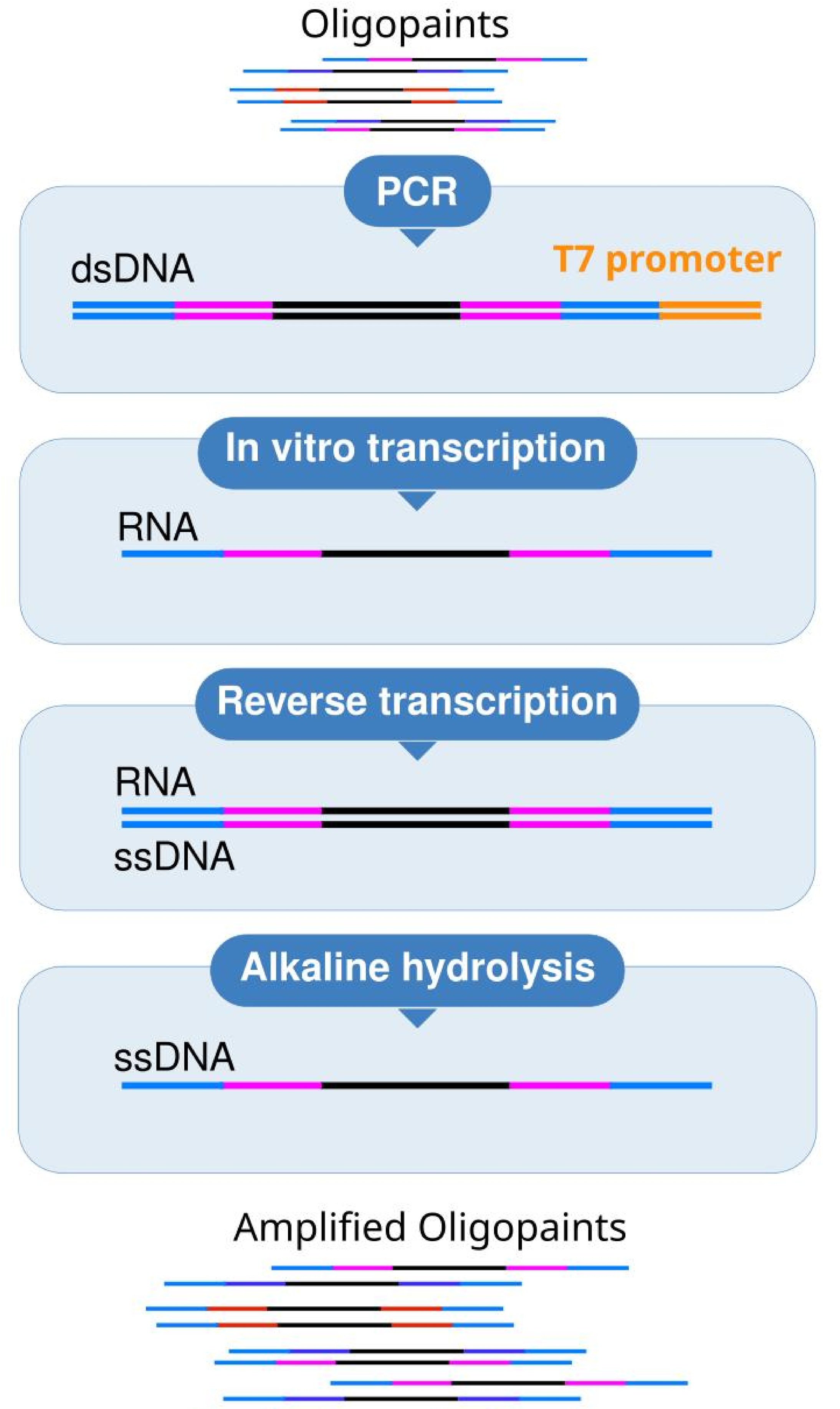

b. Amplification of Oligopaints
Oligopaints
PCR
dsDNA
T7 promoter
In vitro transcription
RNA
Reverse transcription
RNA
ssDNA
Alkaline hydrolysis
ssDNA
Amplified Oligopaints

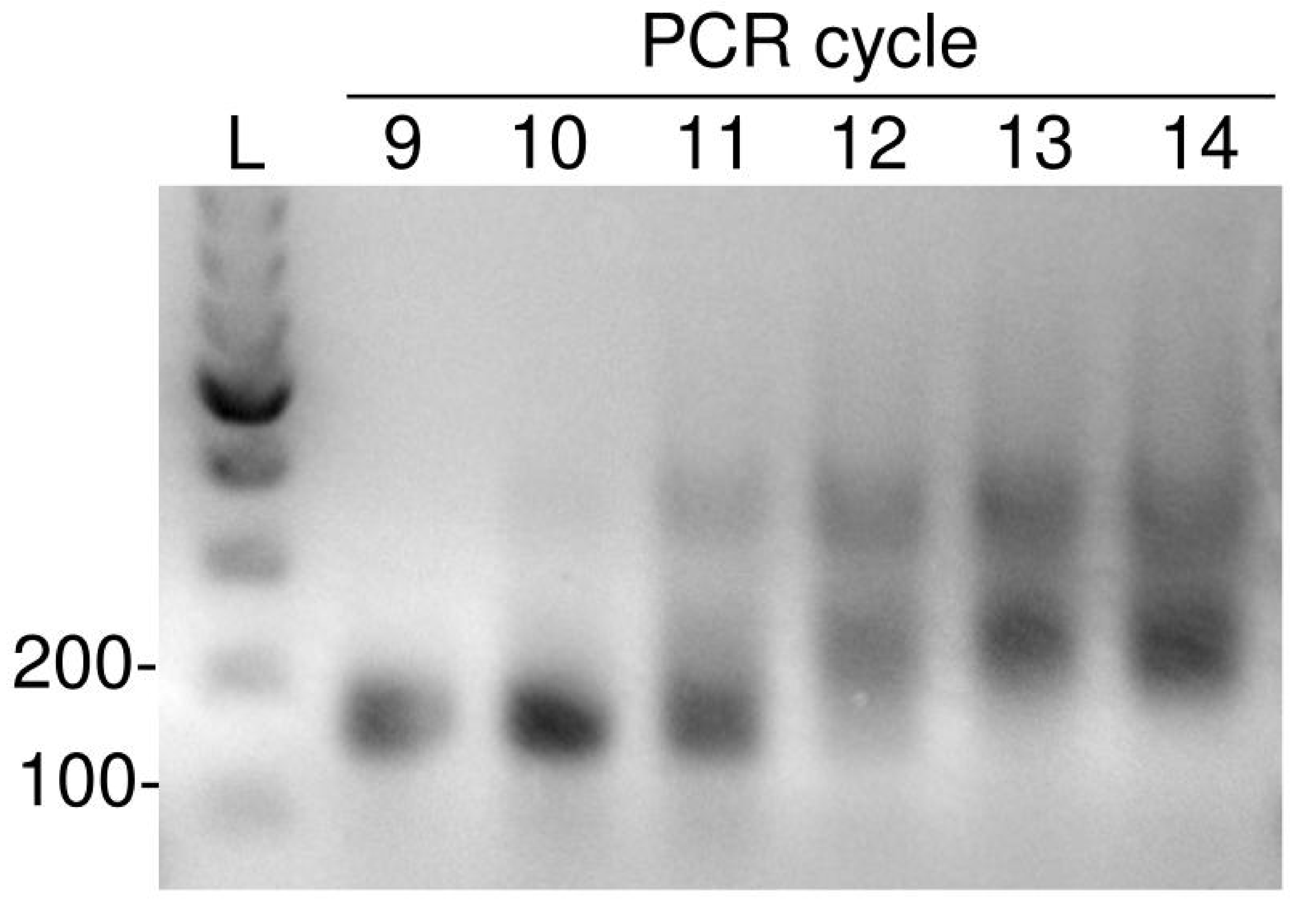

c
Small scale limited-cycle PCR
PCR cycle
L
9
10
11
12
13
14
200-
100-

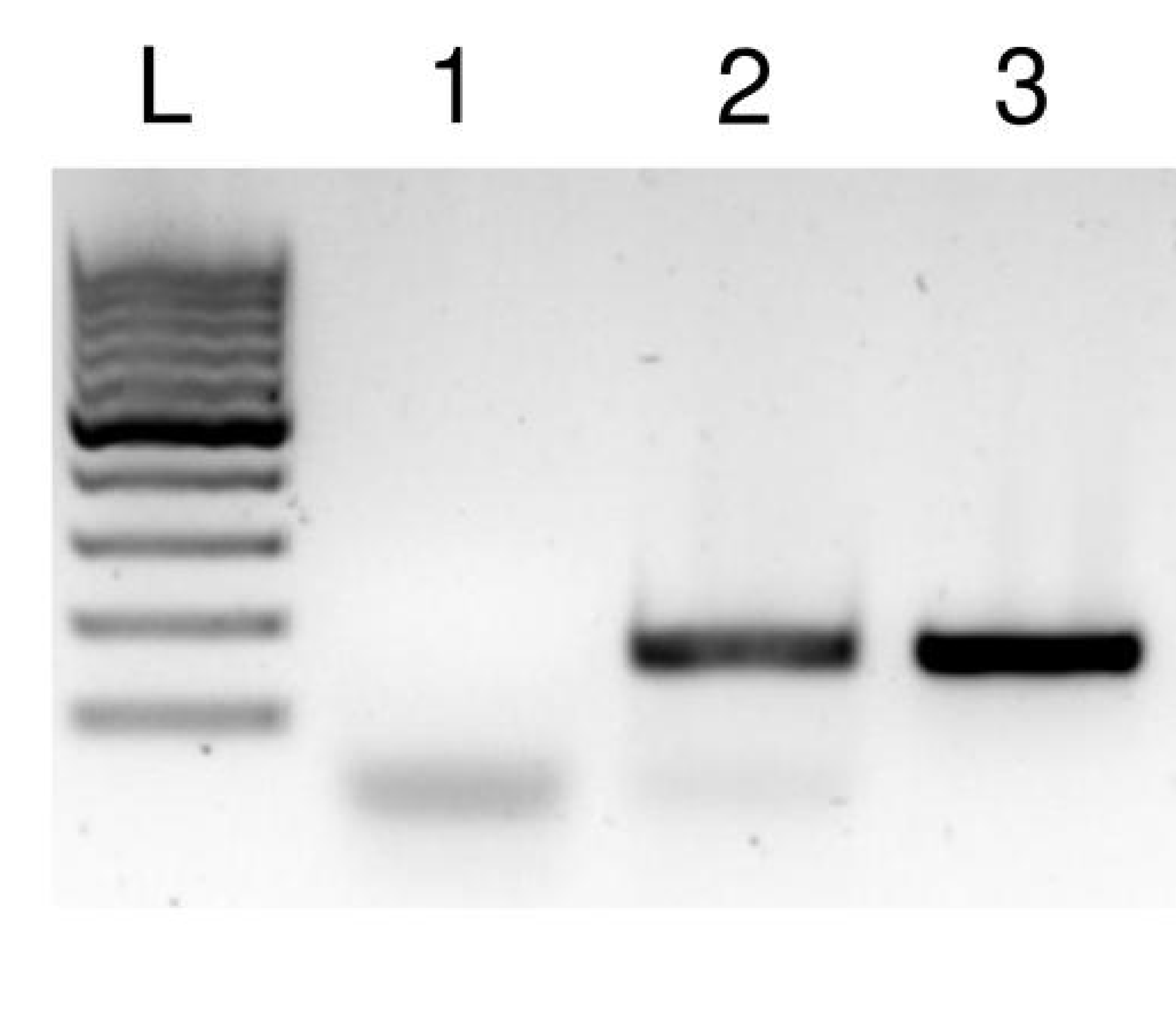

d
Large scale
limited-cycle PCR
L
1
2
3

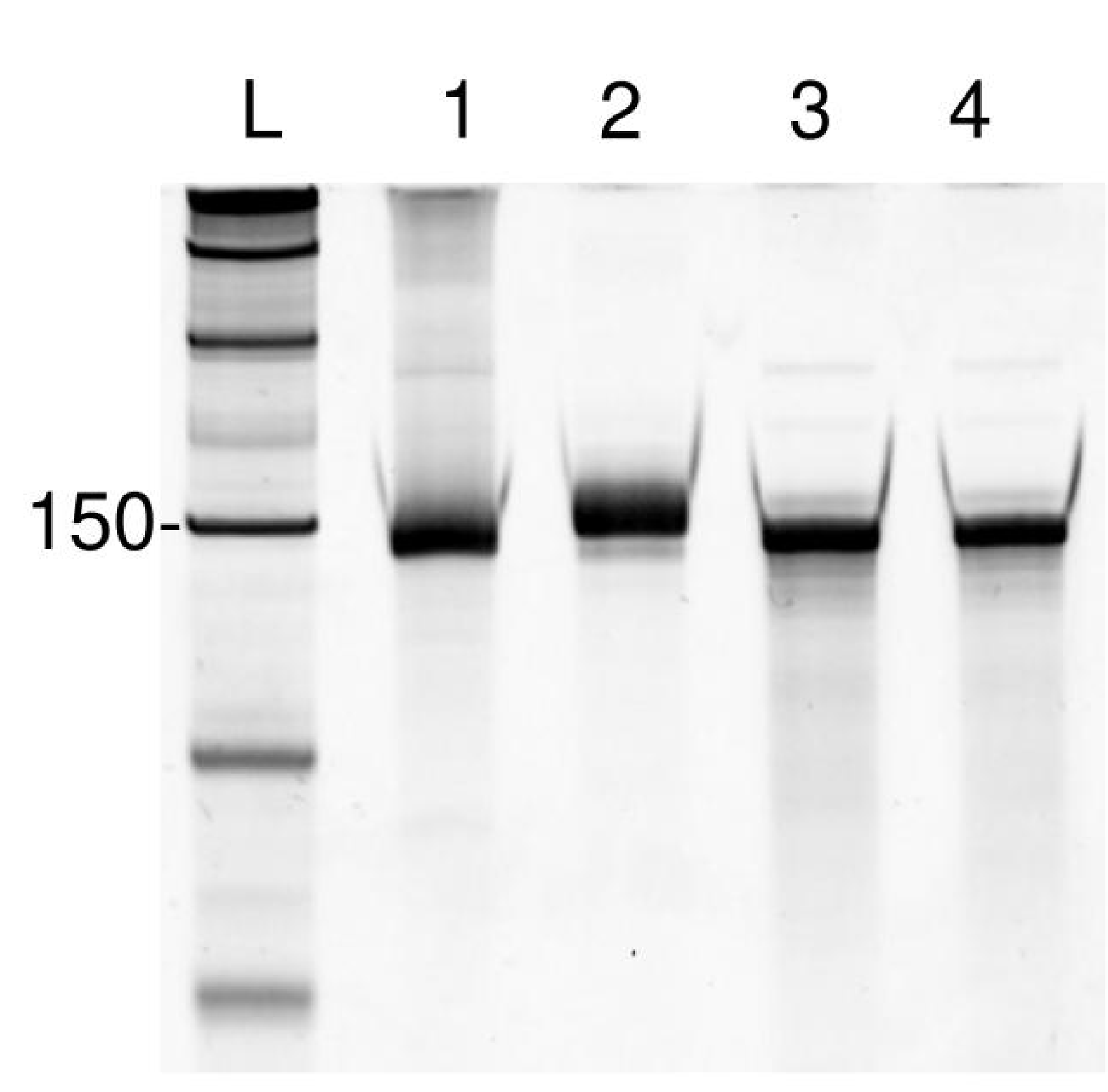

e
Urea PAGE
L
1
2
3
4
150-

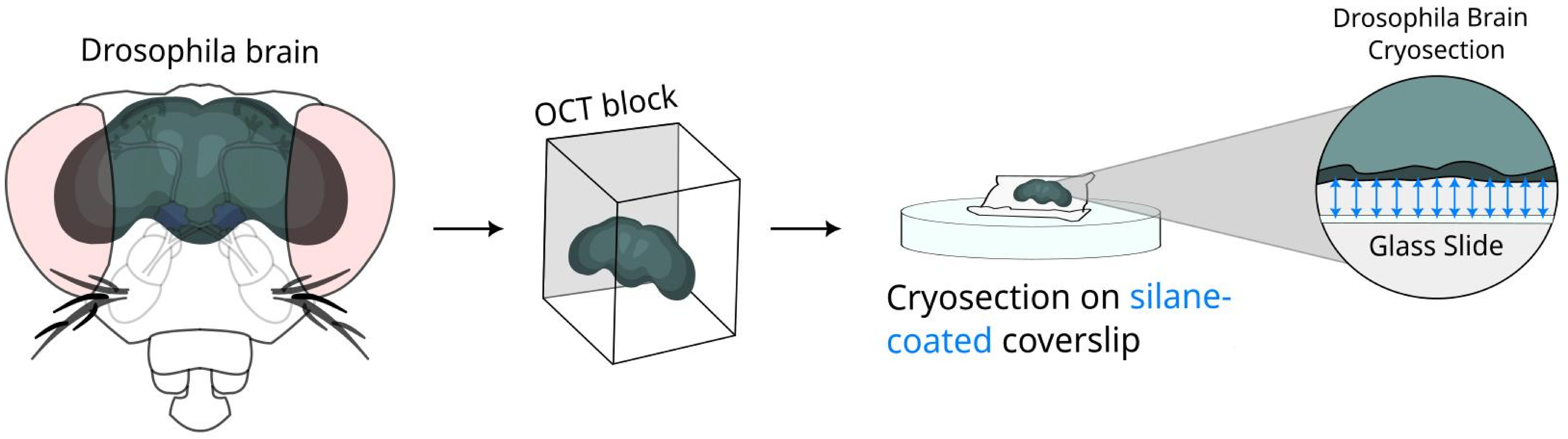

f. Drosophila brain preparation
Drosophila brain
OCT block
Cryosection on silane-
coated coverslip
Drosophila Brain
Cryosection
Glass Slide

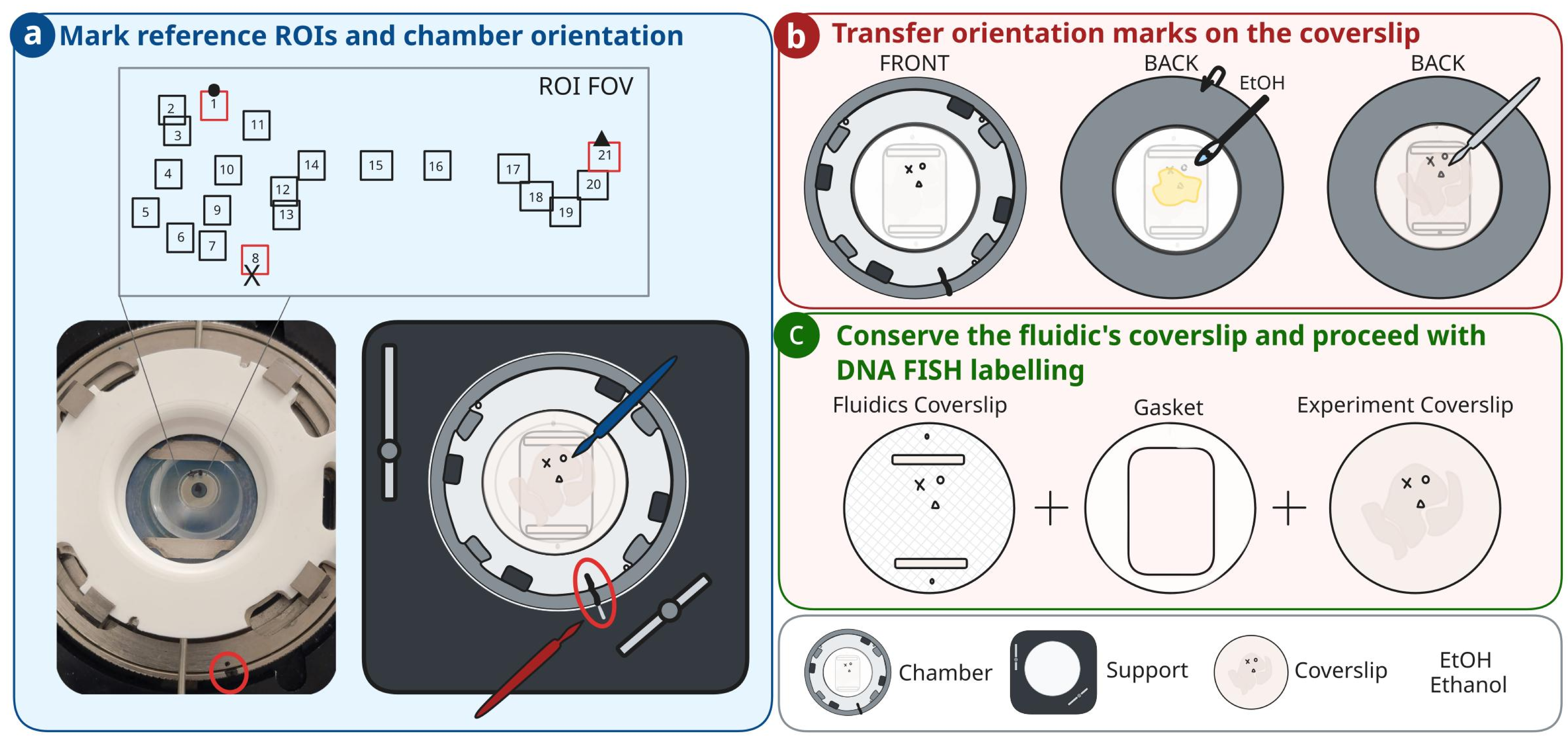

a
Mark reference ROIs and chamber orientation
ROI FOV
1
2
3
4
5
6
7
8
9
10
11
12
13
14
15
16
17
18
19
20
21
b
Transfer orientation marks on the coverslip
FRONT
BACK
EtOH
BACK
C
Conserve the fluidic's coverslip and proceed with DNA FISH labelling
Fluidics Coverslip
Gasket
Experiment Coverslip
Chamber
Support
Coverslip
EtOH
Ethanol

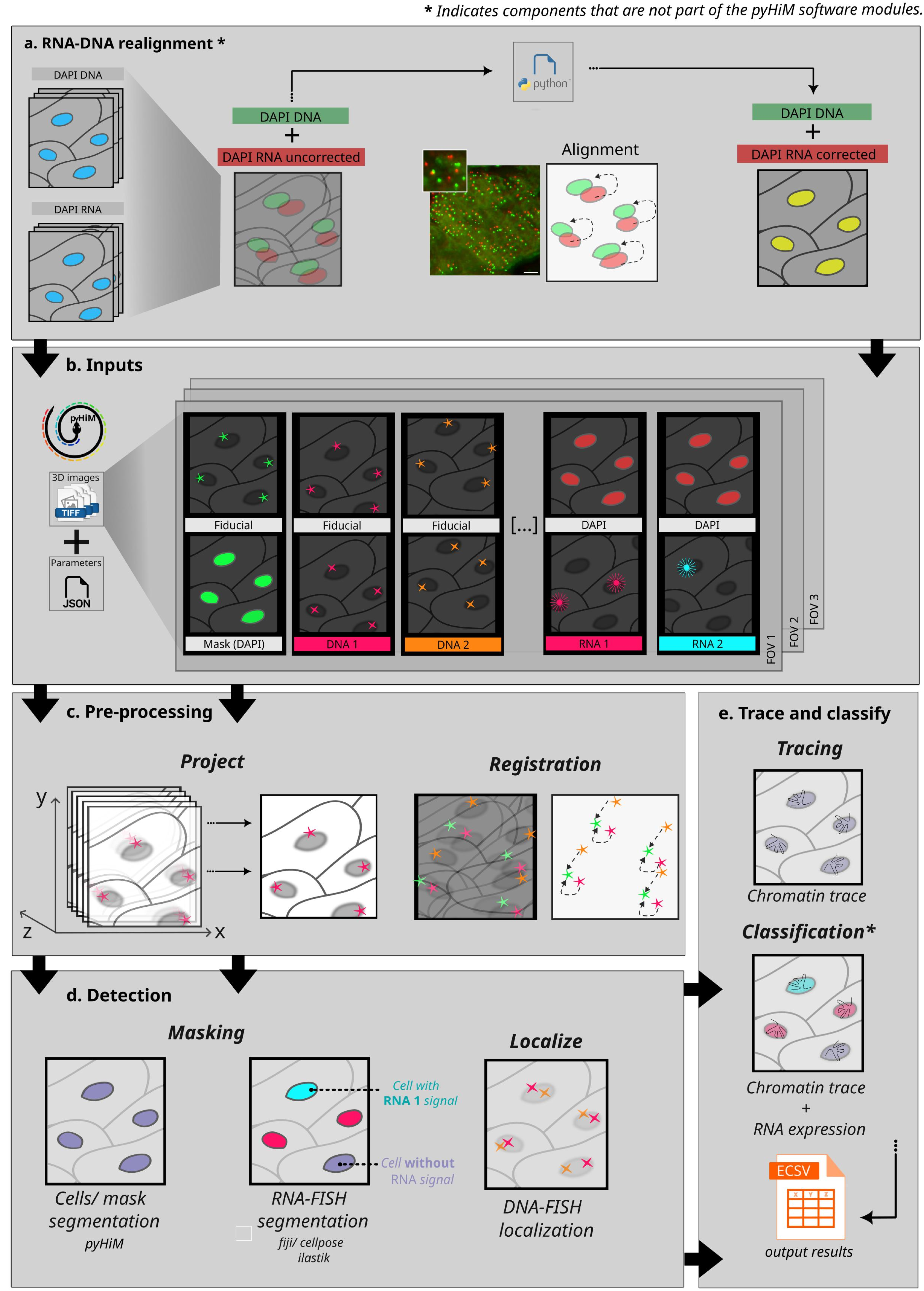
* Indicates components that are not part of the pyHiM software modules.
a. RNA-DNA realignment *
DAPI DNA
DAPI RNA
DAPI DNA
+
DAPI RNA uncorrected
python
Alignment
DAPI DNA
+
DAPI RNA corrected
b. Inputs
pyHiM
3D images
TIFF
+
Parameters
JSON
Fiducial
Fiducial
Fiducial
[...]
DAPI
DAPI
Mask (DAPI)
DNA 1
DNA 2
RNA 1
RNA 2
FOV 1
FOV 2
FOV 3
c. Pre-processing
Project
Registration
y
z
x
d. Detection
Masking
Localize
Cell with
RNA 1 signal
Cell without
RNA signal
Cells/ mask
segmentation
pyHiM
RNA-FISH
segmentation
fiji/ cellpose
ilastik
DNA-FISH
localization
e. Trace and classify
Tracing
Chromatin trace
Classification*
Chromatin trace
+
RNA expression
ECSV
output results